\documentclass[10pt,twoside,notitlepage,twocolumn]{article}
\usepackage{letter}

\newcommand\email[1]{%
	\begingroup
	\renewcommand\thefootnote{}\footnote{#1}%
	\addtocounter{footnote}{-1}%
	\endgroup
}
\begin{document}
\allsectionsfont{\sffamily}

%
%

\title{Osmotic forces modify lipid membrane fluctuations}

\shorttitle{Osmotic forces modify lipid membrane fluctuations}

\author[ ]{Amaresh Sahu$^\ddag$\vspace{-4pt}}

\shortauthor{A.\ Sahu}

\affil[ ]{\hspace{-3pt}McKetta Department of Chemical Engineering, University of Texas, Austin TX 78712, USA}

\date{21 February 2026}

%
%

\twocolumn[
	\begin{@twocolumnfalse}
		\maketitle
		\vspace{-2pt}
		\begin{abstract}
			\noindent\textsf{\textbf{Abstract.}}
In hydrodynamic descriptions of lipid bilayers, the membrane is often
approximated as being impermeable to the surrounding, solute-containing fluid.
However, biological and in vitro lipid membranes are influenced by their
permeability and the resultant osmotic forces---whose effects remain poorly
understood.
Here, we study the dynamics of a fluctuating, planar lipid membrane that is
ideally selective: fluid can pass through it, while solutes cannot.
We find that the canonical membrane relaxation mode, in which internal membrane
forces are balanced by fluid drag, no longer exists over all wavenumbers.
Rather, this mode only exists when it is slower than solute diffusion---%
corresponding to a finite range of wavenumbers.
The well-known equipartition result quantifying the size of membrane undulations
due to thermal perturbations is consequently limited in its validity to the
aforementioned range.
Moreover, this range shrinks as the membrane surface tension is increased, and
above a critical tension the membrane mode vanishes.
Our findings are relevant when interpreting experimental measurements of
membrane fluctuations, especially in vesicles at moderate to high tensions.

		\end{abstract}
		\vskip 1.9em
	\end{@twocolumnfalse}
]
\thispagestyle{empty}

\email{%
	$^\ddag \,$\href{mailto:asahu@che.utexas.edu}{\texttt{asahu@che.utexas.edu}}%
}

\normalsize
\vspace{-15pt}

%
%

\section*{Introduction}

Biological membranes are unique materials that surround the cell, and
compartmentalize its internal organelles.
While such membranes are slowly permeable to water \cite{olbrich-bpj-2000},
large neutral molecules (such as sugars) cannot easily pass through them
\cite{wood-bba-1968}.
The semipermeability of lipid membranes is relevant to both biological and in
vitro scenarios---as in both cases, membranes are essentially always surrounded
by solutes.
While the general equations governing the coupled membrane, fluid, and solute
dynamics were recently obtained \cite{alkadri-pre-2025}, there is not yet a
systematic characterization of membrane behavior.
As a result, the effects of permeability and osmosis on lipid membrane dynamics
remain poorly understood.


The present study investigates how planar lipid membranes act on, and are
influenced by, their surrounding environment.
Flat membranes are found in various biological settings, including the
endoplasmic reticulum \cite{terasaki-cell-2013}, Golgi complex
\cite{tanaka-jemt-1991}, and neuromuscular junction \cite{watanabe-elife-2013}.
In addition, theories of impermeable bilayer fluctuations about a flat surface
are often used to describe the thermal undulations of giant unilamellar vesicles
(GUVs).
We begin by investigating an impermeable membrane, where the dynamics of the
solute are not considered.
By starting with a well-understood system (see e.g.\ Refs.\
\cite{prost-epl-1996, granek-jpf-1997, brown-bpj-2003, chen-prl-2004,
gov-prl-2004, lin-prl-2004, lacoste-epl-2005, lomholt-pre-2006, ben-prl-2011,
loubet-pre-2012, alert-bpj-2015, sapp-pre-2016, turlier}),
we are better equipped to study our primary system of interest: the
semipermeable membrane.
The temporal evolution of the normal modes is solved for within the linear
theory, and the relative time scale of membrane and solute relaxation is found
to dictate the nature of the coupled dynamics.
When solutes relax more quickly, the dynamics are indistinguishable from the
impermeable scenario.
It is then straightforward to present an effective Langevin equation including
thermal perturbations, and obtain the magnitude of the resultant height
fluctuations.
In contrast, when the membrane relaxes more quickly than the surrounding
solutes, the resultant dynamics are qualitatively different from their
impermeable counterpart.
As a result, the character of thermal undulations---long assumed to be that of
an impermeable membrane when analyzing fluctuating GUVs---is no longer known.
Experimental consequences of our findings are highlighted, and avenues for
future work are discussed.

%
%

\section*{The impermeable membrane}

Consider a nearly planar lipid membrane surrounded by water.
For the present analysis, we neglect solutes and make the standard approximation
that the membrane is impermeable to fluid.
Water is characterized by its three-dimensional mass density $ \rhof $, shear
viscosity $ \muf $, and kinematic viscosity
$ \nuf := \muf / \rhof $.
Relevant membrane parameters are the patch length $ \lc $, two-dimensional (2D)
areal density $ \rhom $, surface tension $ \lambdac $, and bending modulus
$ \kb $ \cite{note-bending}.
It is important to note that the surface tension is not a membrane property, but
rather takes the requisite value to enforce areal incompressibility
\cite{steigmann-arma-1999, sahu-thesis}.
As a result, the membrane surface tension can span a wide range of values, and
is often tuned in experiments \cite{gracia-sm-2010, ayala-nc-2023}.
Characteristic values of all parameters are provided in Table \ref{tab_params}.

Our base state is an unperturbed membrane sheet lying in the $ x $--$ y $
plane, surrounded on both sides by a semi-infinite stationary fluid.
We seek to understand how the coupled membrane and fluid system responds to
small perturbations.
To this end, we determine the linearized dynamics of the coupled membrane and
fluid system about the base configuration.
In what follows, an abbreviated analysis of the governing equations is
presented; see \S2 of the Supplemental Material (SM) \cite{supplemental} for
further details.

%
%

\subsection*{The governing equations}

The position $ \bmx $ of the perturbed membrane is given by
$ \bmx (x, y, t) = x \mk \bme_x + y \mk \bme_y + h (x, y, t) \mk \bme_z $,
where
$ x, y \in [0, \lc] $
and
$ \lvert h \rvert \ll \lc $
by construction.
For planar geometries with a stationary base state, the in-plane and
out-of-plane membrane equations are decoupled \cite{sahu-pre-2017,
sahu-pre-2020}.
To solve for the time evolution of the membrane height in the absence of thermal
fluctuations, we require only the $ z $-component of the linear momentum balance
of the membrane---often called the shape equation and given by
\cite{sahu-pre-2020}
\begin{equation} \label{eq_imp_shape}
	\rhom \mk h_{\mkn , \mk t t}
	\, = \, \lambdac \mk h_{\mkn , \mk \alpha \alpha}
	\, - \, \dfrac{\kb}{2} \mk h_{\mkn , \mk \alpha \alpha \beta \beta}
	\, - \, \jjp
	~.
\end{equation}
In Eq.\ \eqref{eq_imp_shape},
$ \alpha, \beta \in {1, 2} $
denote directions in the $ x $--$ y $ plane, with repeated indices summed over
such that (for example)
$ h_{\mkn , \mk \alpha \alpha} = h_{\mkn , \mk x x} + h_{\mkn , \mk y y} $.
Here
$ \pdp_{, t} := \partial \pdp / \partial t $
is the partial time derivative and
$ \jjp := p^+ - p^- $
is the difference in the fluid pressure above ($+$) and below ($-$) the
membrane.
The pressure and velocity fields, $ p^\pm $ and $ v^\pm_{\mkn j} $ (where Roman
indices span $ \{ 1, 2, 3 \} $ and repeated indices are summed over), satisfy
the incompressible Navier--Stokes equations linearized about the base state:
$ v^\pm_{\mkn j, \mk j} = 0 $
and
$
	\rhof \mk v^\pm_{\mkn j, \mk t}
	= \muf \mk v^\pm_{\mkn j, \mk k k}
	- p^\pm_{\mkn , \mk j}
$.
With the no-slip boundary condition
$ h_{\mkn , \mk t} = v^\pm_{\mkn z} $
at the membrane surface, water cannot flow through the membrane and we have a
well-posed set of equations for the coupled system \cite[\S2]{supplemental}.

At this point, we follow standard approaches to solve for the dynamics of the
system.
The membrane height is decomposed into planar Fourier modes as
\begin{equation} \label{eq_imp_height_fourier}
	h (x, y, t)
	\, = \, \sum_{\bmq}
		\hath \,
		e^{i \mk (\qx x \, + \, \qy y) } \,
		e^{- \omegaq t}
	~,
\end{equation}
where the wavevector
$ \bmq = (\qx, \qy) $
has corresponding wavenumber
$
	q
	:= (q_x^{\, 2} + q_y^{\, 2})^{1/2}
$
$
	\sim 10^{-4} \text{--} 10^{-1} \text{ nm}^{-1} \!\!
$,~~%
$ \omegaq $ is the associated relaxation frequency, and normal modes are assumed
independent.
Fluid unknowns are similarly decomposed into planar Fourier modes, with the
$ z $-dependence to be solved for.
The fluid pressure, for example, is expressed as
\begin{equation} \label{eq_imp_pressure_fourier}
	p^\pm (x, y, z, t)
	\, = \, \sum_{\bmq}
		\hatppm (z) \,
		e^{i \mk (\qx x \, + \, \qy y) } \,
		e^{- \omegaq t}
	~.
\end{equation}
Our main task is to solve for all frequencies $ \omegaq $ satisfying the
governing equations.
In doing so, we also solve for the pressure and velocity fields in the fluid.

\begin{table}[t!]
	\centering%
	\small
	\setlength{\tabcolsep}{2pt}
	\renewcommand{\arraystretch}{1.10}
	\caption{%
		Membrane, fluid, and solute parameters.%
	}%
	\vspace{-2pt}
	\begin{tabular}{l c c c}
		\hline
		\hline
		Parameter & Sym.\ & Value & Ref. \\
		\hline
		Patch size &
		$\lc$ &
		$ 10^2\text{--}10^3 $ nm &
		\cite{terasaki-cell-2013, watanabe-elife-2013} \\
		Density &
		$ \rhom $ &
		$ 10^{-8}$ pg/nm$^2 $ &
		\cite{parkkila-langmuir-2018} \\
		Bending modulus &
		$ \kb $ &
		$ 10^2~\text{pN} \! \cdot \! \text{nm} $ &
		\cite{pecreaux-epje-2004} \\
		Surface tension &
		$ \lambdac $ &
		$ 10^{-4}$--$10^{-1} $ pN/nm &
		\cite{pecreaux-epje-2004, dai-jn-1998} \\
		Impermeability &
		$ \kappa $ &
		$ 10^3\text{--}10^5 $ pN$\cdot$\textmu sec/nm$^3 $ &
		\cite{olbrich-bpj-2000, ipsen-ch14} \\
		\arrayrulecolor{black!30}\hline
		Density &
		$ \rhof $ &
		$ 10^{-9}$ pg/nm$^3 $ &
		-- \\
		Shear viscosity &
		$ \muf $ &
		$ 10^{-3} $ pN$\cdot$\textmu sec/nm$ ^2 $ &
		-- \\
		Kinematic viscosity &
		$ \nuf $ &
		$ 10^{6} $ nm$^2 $\!/\textmu sec &
		-- \\
		\hline
		Concentration &
		$ \cz $ &
		$ 10^{-1}$ nm$^{-3} $ &
		\cite{dimova-jpcm-2006} \\
		Diffusivity &
		$ D $ &
		$ 10^{3}$ nm$^2 $\!/\textmu sec &
		\cite{gosting-jacs-1949} \\
		\arrayrulecolor{black}\hline
		\hline
	\end{tabular}
	\label{tab_params}
\end{table}

%
%

\subsection*{The frequency solutions}

\begin{figure}[!b]
	\centering
	\includegraphics[width=0.80\columnwidth]{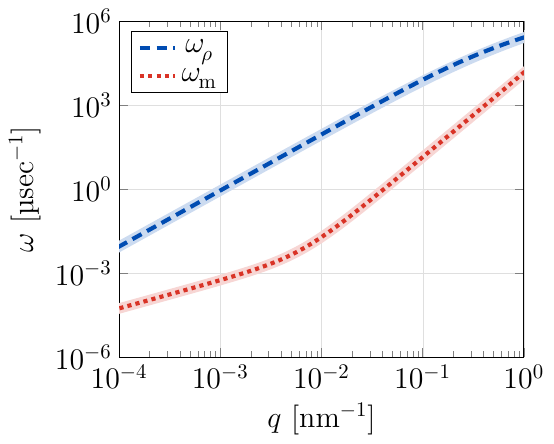}
	\caption{%
		Plot of the two impermeable frequency solutions as a function of wavenumber.
		Blue dashed and red dotted lines are respectively the exact inertial and
		membrane branches, obtained numerically.
		The approximate solutions in Eq.\ \eqref{eq_imp_tilomega} are shown as
		thick, transparent bands.
		Relevant parameters are
		$ \rhof = 10^{-9} $ pg$/$nm$^3$,
		$ \nuf  = 9 \! \cdot \! 10^{-5} $ nm$^2$/\textmu sec,
		$ \rhom = 10^{-8} $ pg$/$nm$^2$,
		$ \lambdac = 2 \! \cdot \! 10^{-3} $ pN$/$nm,
		and
		$ \kb = 10^2 $ pN$\cdot$nm.
	}
	\label{fig_imp_omega}
\end{figure}

As we limit ourselves to the linearized dynamics, all unknowns of a particular
wavevector $ \bmq $ are proportional to the membrane height $ \hath $.
We assume
$ \hath \ne 0 $
and combine all equations governing membrane and fluid into a single equation
for $ \omegaq $, namely the dispersion relation (see \S2.3 of the SM
\cite{supplemental}).
Given the physically-motivated requirement that all fluid perturbations decay
far from the membrane surface, there are two frequencies at each wavenumber,
which are plotted in Fig.~\ref{fig_imp_omega}.
The aforementioned equation for $ \omegaq $ is not straightforward to interpret%
---though we find the dynamics are well-approximated by
\begin{equation} \label{eq_imp_quadratic}
	\rhoeff \, \omegaq^{\, 2}
	\, - \, 4 \mk \muf \mk q \, \omegaq
	\, + \, E
	\, = \, 0
	~.
\end{equation}
Equation \eqref{eq_imp_quadratic}, with the temporal ansatz
$ h \sim e^{- \omega t} $
in Eq.\ \eqref{eq_imp_height_fourier},
reveals the dynamics of the membrane are analogous to that of the spring--mass%
--damper from introductory mechanics.
Here
$ \rhoeff := \rhom + \mk 4 \mk \rhof / q $
$ \sim 10^{-7} \text{--} 10^{-5} \, \text{pg} / \text{nm}^2 $
approximates the combined inertia of membrane and fluid,
$ 4 \muf \mk q $
$
	\sim 10^{-7} \text{--} 10^{-4} \,
	\text{pN$ \cdot $\textmu sec} / \text{nm}^3
$
is the well-known hydrodynamic drag from the surrounding fluid, and
$ E := \lambdac \mk q^2 + \tfrac{1}{2} \mk \kb \mk q^4 $
$
	\sim 10^{-12} \text{--} 10^{-2} \, \text{pN}/\text{nm}^3
$
captures the energetics of small membrane fluctuations:
$ \tfrac{1}{2} \, \lc^{\, 2} \mk E \mk \lvert \hath^{\, 2} \rvert $
is the energy of mode $ \bmq $ to quadratic order \cite{brochard-jp-1975}.
As the system is overdamped,
$
	4 \muf \mk q \mkn \gg \! \sqrt{\rhoeff \mk E}
$
and the two solutions to Eq.\ \eqref{eq_imp_quadratic} are well-approximated by
\begin{equation} \label{eq_imp_tilomega}
	\tilomegarho
	\, := \, \dfrac{4 \mk \muf \mk q}{\rhoeff}
	\qqandqq
	\tilomegam
	\, := \, \dfrac{E}{4 \mk \muf \mk q}
	~.
\end{equation}
In Eq.\ \eqref{eq_imp_tilomega}, the `tilde' accent indicates that Eq.\
\eqref{eq_imp_quadratic} is only an approximation of the true dynamics; the
subscripts $ \rho $ and $ \mathrm{m} $ designate the inertial and membrane
branches.
As shown in Fig.\ \ref{fig_imp_omega}, the frequencies in Eq.\
\eqref{eq_imp_tilomega} are a good approximation to the true solution.

%
%
\medskip
\noindent\textbf{\textsf{The membrane response.}}
Following Eqs.\ \eqref{eq_imp_quadratic} and \eqref{eq_imp_tilomega}, let us
denote the two exact frequency branches as $ \omegarho $ and $ \omegam $.
The membrane height, in real space, is then given by [cf.\ Eq.\
\eqref{eq_imp_height_fourier}]
\begin{equation} \label{eq_imp_height_branches}
	h (x, y, t)
	\, = \, \sum_{\bmq} \! \Big(
			\hathrho \mk e^{- \omegarho t}
			\mk + \mk \hathm \mk e^{- \omegam t}
		\Big)
		\mk e^{i \mk (\qx x \, + \, \qy y) }
		,
\end{equation}
where $ \hathrho $ and $ \hathm $ are determined upon specification of the
initial conditions: $ h $ and $ \partial h / \partial t $ at time
$ t = 0 $.
Since
$ \omegarho \gg \omegam $
over all wavenumbers, the inertial branch quickly decays and the dynamics of the
coupled system are captured by the membrane branch.
Prior studies are accordingly justified in neglecting inertial effects
throughout their investigations of planar lipid membrane behavior
\cite{prost-epl-1996, granek-jpf-1997, brown-bpj-2003, chen-prl-2004,
gov-prl-2004, lin-prl-2004, lacoste-epl-2005, lomholt-pre-2006, ben-prl-2011,
loubet-pre-2012, alert-bpj-2015, sapp-pre-2016, turlier, camley-prl-2010,
oppenheimer-bpj-2009, stone-jfm-2015, camley-jcp-2015, suja-prf-2025}.

%
%
\medskip
\noindent\textbf{\textsf{The fluid response.}}
To understand the dynamics of the fluid as the membrane relaxes, consider two
disturbances to a specific mode $ \bmq $.
In the first case,
$ \hat{h}^{\mathrm{m}}_{\bmq} = 0 $
and only the inertial mode is excited, while in the second case
$ \hat{h}^\rho_{\bmq} = 0 $
and only the membrane mode is excited
[see Eq.\ \eqref{eq_imp_height_branches}].
In both cases, the pressure field \eqref{eq_imp_pressure_fourier} is given by
$
	\hatppm
	= \mp \tfrac{1}{2} (
		\rhom \mk \omegaq^2
		+ E
	) \, \hath \mk e^{- q \lvert z \rvert}
$.
The magnitude of the pressure at $ z = 0 $ determined by Eq.\
\eqref{eq_imp_shape}, and the decay length $ 1 / q $ is a consequence of the
pressure field being harmonic:
$ p^\pm_{\mkn , \mk j j} = 0 $.
In contrast, fluid velocities decay over the two length scales $ 1 / q $ and
$ 1 / \knu $, where
$ \knu := ( q^2 - \omega / \nuf )^{1/2} $
is a modified wavenumber that accounts for the inertia of the fluid (see \S94 of
Ref.\ \cite{chandrasekhar} and \S2.1 of the SM \cite{supplemental}).
As a consequence of the time scale separation
$ \omegarho \gg \omegam $,
$ 1 / \barq \approx 1 / \knu $
on the membrane branch and
$ 1 / \barq \ll 1 / \knu $
on the inertial branch.

\begin{figure*}[t!]
	\centering
	\begin{subfigure}[b]{0.32\textwidth}
		\centering
		\includegraphics[width=\textwidth]{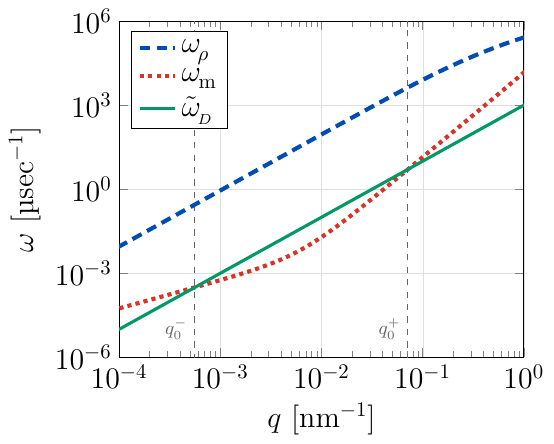}
		\caption{impermeable overlay}
		\label{fig_semi_diff_omega}
	\end{subfigure}
	\begin{subfigure}[b]{0.32\textwidth}
		\centering
		\includegraphics[width=\textwidth]{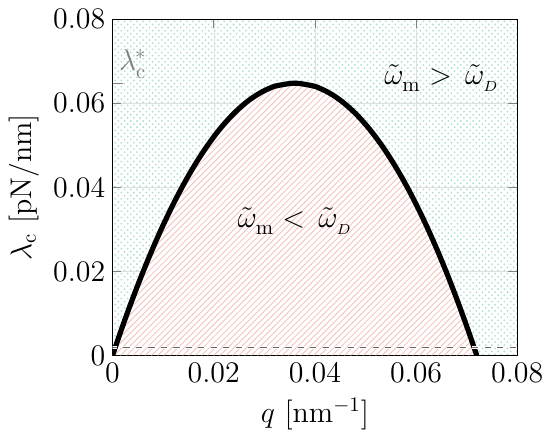}
		\caption{dome comparing frequencies}
		\label{fig_semi_diff_dome}
	\end{subfigure}
	\begin{subfigure}[b]{0.32\textwidth}
		\centering
		\includegraphics[width=\textwidth]{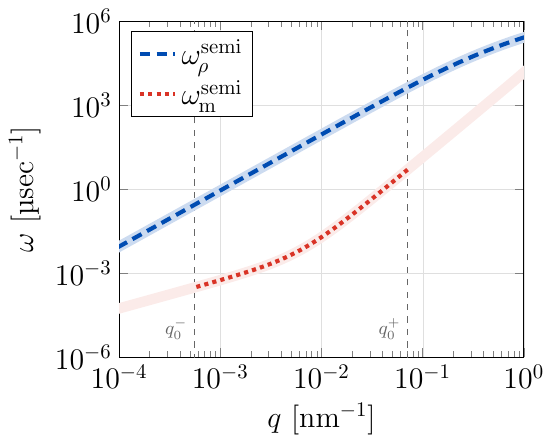}
		\caption{semipermeable branches}
		\label{fig_semi_diff_sols}
	\end{subfigure}
	\caption{%
		Relevance of the diffusive time scale to the dynamics of the semipermeable
		system.
		(a) An overlay of $ \tilomegad $ (solid green line) on top of the
		impermeable frequencies in Fig.\ \ref{fig_imp_omega}.
		There are two wavenumbers for which the impermeable membrane frequency
		$ \omegamimp $ equals $ \tilomegad $; approximating the former as
		$ \tilomegam $ yields the two crossover wavenumbers $ \qzm $ and $ \qzp $
		\eqref{eq_semi_qzpm}---shown as vertical, dashed lines.
		(b) Plot of $ \qzpm $ over a range of the base membrane tension
		$ \lambdac $, which varies across systems.
		The surface tension from (a) of
		$ \lambdac = 2 \! \cdot \! 10^{-3} $ pN$/$nm
		is shown as the horizontal dashed line.
		There exists a critical tension $ \lambdacast $ \eqref{eq_semi_lambdacast}
		above which
		$ \tilomegam > \tilomegad $
		for all wavenumbers.
		(c) Plot of the real parts of the semipermeable frequencies (dotted or
		dashed solid lines), overlaid on top of the impermeable frequencies (thick,
		transparent bands) for comparison.
		The inertial branch is essentially unchanged.
		The membrane branch only exists when
		$ \omegam < \tilomegad $%
		---and is thus confined to the range
		$ q \in (\qzm, \qzp) $.
		For base membrane tensions
		$ \lambdac > \lambdacast $,
		the membrane branch vanishes altogether.
	}
	\label{fig_semi_diff}
\end{figure*}

%
%

\section*{The semipermeable membrane}

With an understanding of the impermeable membrane, we are primed to investigate
the primary system of interest: a semipermeable membrane, in which fluid flows
through the membrane while solutes dissolved in the fluid do not.
Solutes are characterized by a base concentration
$ \cz \! \sim \! 0.1 $ $ \! \text{nm}^{-3} $
\cite{dimova-jpcm-2006} and diffusion constant
$ D \! \sim \! 10^3 $ $ \! \text{nm}^2 / \text{\textmu sec} $
\cite{gosting-jacs-1949}.
In addition, the membrane impermeability $ \kappa $ relates the flux of water
through the bilayer to its driving force \cite{alkadri-pre-2025}; here
$
	\kappa
	\mkn \sim \mkn 10^3\text{--}10^5
$ $
	\! \text{pN} \! \cdot \! \text{\textmu sec}/\text{nm}^3
$
\cite{olbrich-bpj-2000, ipsen-ch14}.
Dynamical effects from water permeation are of order
$ \muf \mk q / \kappa $:
a dimensionless parameter that we call the permeability number $ \Pm $,
ranging from
$ 10^{-7} \text{--} 10^{-12} $
over wavenumbers
$ q \in ( 10^{-4}, 10^{-1} ) ~ \text{nm}^{-1} $.
We will show
$ \Pm \rightarrow 0 $
in the limit of an impermeable membrane; lipid bilayers are thus nearly
impermeable.

%
%

\subsection*{The governing equations}

The membrane shape equation \eqref{eq_imp_shape} is unchanged in the presence of
solutes \cite{alkadri-pre-2025}.
The perturbed concentrations $ c^\pm $ above and below the membrane, with
$ \lvert c^\pm \rvert \ll \cz $,
evolve according to the diffusion equation
$ c^\pm_{\mkn , \mk t} = D \mk c^{\pm}_{\mkn , \mk j j} $.
Solute impermeability is enforced by requiring the total solute flux through the
membrane---comprised of diffusive and convective contributions---to be zero:
\begin{equation} \label{eq_semi_no_flux}
	- D \mk c^\pm_{\mkn , \mk z}
	\, + \, \cz \mk \big(
		v^\pm_{\mkn z}
		- h_{\mkn , \mk t}
	\big)
	\, = \, 0
	~.
\end{equation}
In Eq.\ \eqref{eq_semi_no_flux}, velocities and concentration gradients are
evaluated at the membrane surface.
The equation governing water permeation through an arbitrarily curved and
deforming lipid membrane was recently obtained by A.M.\ \textsc{Alkadri} and
K.K.\ \textsc{Mandadapu} \cite{alkadri-pre-2025}, who extended several seminal
findings \cite{starling-jp-1896, staverman-rtc-1951, kedem-bba-1958}.
These authors used techniques from irreversible thermodynamics and differential
geometry to show fluid flow through the membrane is driven by differences in
hydrodynamic tractions and osmotic pressures across the membrane surface.
For nearly planar geometries and an ideally selective membrane, the findings
from Ref.\ \cite{alkadri-pre-2025} simplify to
\begin{equation} \label{eq_semi_permeability}
	\kappa \mk \big(
		v^\pm_{\mkn z}
		- h_{\mkn , \mk t}
	\big)
	\, = \, \kBT \mk \jjc
	\, - \, \jjp
	~,
\end{equation}
where $ \kB $ is Boltzmann's constant, $ \vartheta $ is the absolute
temperature, and
$ \jjc := c^+ - c^- $
is the jump in solute concentration at the membrane surface.
In the language of irreversible thermodynamics \cite{prigogine, degroot-mazur},
Eq.\ \eqref{eq_semi_permeability} presents a linear relationship between the
thermodynamic driving force
$ ( \kBT \mk \jjc - \jjp ) $
and the corresponding thermodynamic flux
$ ( v^\pm_{\mkn z} - h_{\mkn , \mk t} ) $,
with the latter a measure of fluid flow through the membrane.
The phenomenological parameter $ \kappa $, understood as a transport coefficient
\cite{alkadri-pre-2025}, captures the resistance to flow: for a given driving
force, a larger value of $ \kappa $ corresponds to a smaller water flux through
the membrane.
Equation \eqref{eq_semi_permeability} is equivalently expressed as
$
	v^\pm_{\mkn z}
	- h_{\mkn , \mk t}
	= (\kBT \mk \jjc
	- \jjp) / \kappa
$;
in the limit of
$ \kappa \rightarrow \infty $
or
$ \Pm \rightarrow 0 $
we find
$ v^\pm_{\mkn z} \rightarrow h_{\mkn , \mk t} $
and recover the impermeable scenario.

%
%

\subsection*{The diffusive time scale}

The presence of solutes in the surrounding fluid introduces a diffusive
frequency
\begin{equation} \label{eq_semi_tilomegad}
	\tilomegad
	\, := \, q^2 D
	~.
\end{equation}
Figure \ref{fig_semi_diff_omega} shows $ \tilomegad $ plotted on top of the
curves in Fig.\ \ref{fig_imp_omega}; note the diffusive and impermeable membrane
frequencies can be comparable.
Let us denote $ \qzpm $ as the two wavenumbers at which the diffusive and
membrane time scales are equal.
From Eqs.\ \eqref{eq_imp_tilomega} and \eqref{eq_semi_tilomegad}, we find
\begin{equation} \label{eq_semi_qzpm}
	\qzpm
	\, = \, \dfrac{4 \mk \muf \mk D}{\kb} \, \bigg[\mk
		1
		\, \pm \, \bigg(
			1
			\, - \, \dfrac{\lambdac \mk \kb}{8 \mk {\muf}^{\! 2} D^2}
		\bigg)^{\!\! 1/2 \,\,}
	\mk\bigg]
	~,
\end{equation}
with
$ \tilomegam < \tilomegad $
when
$ q \in (\qzm, \qzp) $,
and
$ \tilomegam > \tilomegad $
otherwise.
Since the base surface tension
$ \lambdac \sim 10^{-4} \text{--} 10^{-1} $ pN/nm
is not a material parameter, but
rather varies---and can be tuned---across systems, $ \qzpm $ is plotted at
different tensions $ \lambdac $ in Fig.\ \ref{fig_semi_diff_dome} to reveal a
dome-like envelope.
There exists a critical base tension
\begin{equation} \label{eq_semi_lambdacast}
	\lambdacast
	\, := \, \dfrac{8 \mk {\muf}^{\! 2} D^2}{\kb}
	\, \sim \, 6 \cdot 10^{-2} \text{ pN/nm}
\end{equation}
at the top of the dome, above which no real $ \qz $ exists.
We also note that the bilayer tears at the lysis tension
$ \lambdal \sim 1 \text{--} 10 $ pN/nm \cite{olbrich-bpj-2000}.
Thus, when the base membrane tension $ \lambdac $ lies between
$ \lambdacast $ and $ \lambdal $, we have
$ \tilomegam > \tilomegad $
over all wavenumbers.
As we will see, the relative values of $ \tilomegam $ and $ \tilomegad $---and
thus whether a mode is ``inside'' (i.e.\ below) or ``outside'' (i.e.\ above) the
dome in Fig.\ \ref{fig_semi_diff_dome}---govern different regimes of membrane
behavior.

%
%

\subsection*{The frequency solutions}

Membrane and fluid unknowns are expressed as in the impermeable analysis, with
the concentration expanded in planar normal modes as [cf.\ Eq.\
\eqref{eq_imp_pressure_fourier}]
\begin{equation} \label{eq_semi_concentration_fourier}
	c^\pm (x, y, z, t)
	\, = \, \sum_{\bmq}
		\hatcpm (z) \,
		e^{i \mk (\qx x \, + \, \qy y) } \,
		e^{- \omegaq t}
	~.
\end{equation}
In our linearized description, all perturbed bulk quantities are again
proportional to the membrane height.
Assuming a nontrivial solution
($ \hath \ne 0 $),
it is straightforward to arrive at the dispersion relation, as shown in \S3.3 of
the SM \cite{supplemental}.
While $ \omegaq $ cannot be solved for analytically, we take advantage of
the smallness of $ \Pm $: since the membrane is nearly impermeable,
semipermeable frequencies are close to their impermeable counterparts.
In what follows, we investigate how the dynamics of the impermeable branches are
altered by the presence of solutes in the surrounding medium.

%
%
\medskip
\noindent\textbf{\textsf{The inertial branch.}}
We begin by assuming concentration modes evolve with the impermeable inertial
frequency $ \omegarhoimp $, such that for a single mode,
$ \cpm_{, t} = - \omegarhoimp \mk \cpm $
(see Eq.\ \eqref{eq_semi_concentration_fourier}).
As a consequence, the diffusion equation
$ \cpm_{, t} = D \mk \cpm_{, j j} $
for the mode $ \bmq $ simplifies to
$
	- \omegarhoimp \mk \hatcpm
	= D ( -q^2 \mk \hatcpm + \md^2 \hatcpm{} / \md z^2 )
$%
---which after some rearrangement yields
\begin{equation} \label{eq_semi_sol_diff_cpm}
	\ddt{\hatcpm}{z}
	\, = \, - q^2 \mk \tileta^2 \mk \hatcpm
	~.
	\qquad
	\text{Here}
	\qquad
	\tileta
	\, := \mk \bigg(
		\dfrac{\omegarhoimp}{\tilomegad}
		\, - \, 1
	\bigg)^{\!\! 1 \mkn / \mkn 2}
\end{equation}
is a positive, real, dimensionless parameter defined for notational convenience.
Equation \eqref{eq_semi_sol_diff_cpm} admits the two purely oscillatory
solutions
$ \hatcp \sim e^{\pm i \mk q \mk \tileta \mk z } $.
As these standing waves do not decay as
$ z \rightarrow \infty $,
we seek the lowest-order concentration correction due to the semipermeability of
the membrane.%
\footnote{%
	For conciseness, only the concentration field above the membrane is presented;
	see the SM \cite[\S\S$\,$3.2--3.3]{supplemental} for all details.%
}%

We first recognize that water permeation alters the hydrodynamic drag on the
membrane by the surrounding fluid.
Equations \eqref{eq_semi_no_flux} and \eqref{eq_semi_permeability} together
yield
$ \jjp = \kBT \mk \jjc - (\kappa \mk D / \cz) \mk c^\pm_{\mkn , \mk z} \mk $,
which---along with Eq.\ \eqref{eq_imp_shape}---shows standing waves in the
concentration add a hydrodynamic drag force 90$^\circ$ out of phase with the
membrane velocity.
We show in the SM \cite[\S3.3.2]{supplemental} that an approximate dispersion
relation for the semipermeable membrane is given by [cf.\ Eq.\
\eqref{eq_imp_quadratic}]
\begin{equation} \label{eq_semi_quadratic}
	\rhoeff \, \omegaq^{\, 2}
	\, - \, 4 \mk \muf \mk q \, \omegaq
	\, + \, E
	\, = \, \pm \, \dfrac{
		8 \mk i \mk \Pm \mk \Os
	}{
		\tileta
	} \, ( 4 \mk \muf \mk q ) \, \omegaq
	~.
\end{equation}
In Eq.\ \eqref{eq_semi_quadratic},
$
	\Os
	:= \kBT \mk \cz / (D \mk \kappa \mk q)
$
$
	\sim 10^{-8} \text{--} 10^{-3}
	\ll 1
$
is a dimensionless parameter that arises when Eqs.\ \eqref{eq_semi_no_flux} and
\eqref{eq_semi_permeability} are combined.
The solution of Eq.\ \eqref{eq_semi_quadratic} closest to $ \omegarhoimp $, to
lowest order in the product $ \Pm \mk \Os $ of small parameters, is given by
\begin{equation} \label{eq_semi_sol_inertia_omega}
	\omegarho
	\, \approx \, \omegarhoimp \, \big(
		1
		\, \pm \, 8 \mk i \mk \Pm \mk \Os / \tileta
	\mk\big)
	~.
\end{equation}
Membrane permeability and osmotic forces accordingly introduce exceedingly slow
temporal oscillations to the inertial dynamics of the coupled system.

The modification to the inertial frequency in Eq.\
\eqref{eq_semi_sol_inertia_omega} also alters the spatial concentration of the
solutes.
Substituting Eq.\ \eqref{eq_semi_sol_inertia_omega} into the diffusion equation
and solving for the concentration field yields
\begin{align}
	\hatcp
	\, &\sim \, \exp \mk \bigg\{
		\pm \, i \mk q \mk \tileta \mk z
		\, - \, \dfrac{4 \mk \Pm \mk \Os \mk q \mk z}{1 - \tilomegad / \omegarhoimp}
	\bigg\}
	\label{eq_semi_sol_inertia_hatcp}
	~.
\end{align}
Importantly,
$ \hatcp \rightarrow 0 $
as
$ z \rightarrow \infty $
and concentration perturbations due to the fluctuating membrane decay far from
the membrane surface.
We have thus found a physically meaningful solution for both the frequency
\eqref{eq_semi_sol_inertia_omega} and concentration
\eqref{eq_semi_sol_inertia_hatcp} over all wavenumbers.
The validity of our approximate analysis is confirmed in Fig.\
\ref{fig_semi_diff_sols}: the impermeable branch is visually indistinguishable
from the real portion of its semipermeable counterpart.

%
%
\medskip
\noindent\textbf{\textsf{The membrane branch when diffusion is slow.}}
As in the analysis of the inertial branch, we start by assuming the solute
concentration evolves at the frequency of the impermeable system---in this case,
the membrane frequency $ \omegamimp $.
For wavenumbers
$ q \notin (\qzm, \qzp) $,
diffusion is slow relative to the membrane and
$ \omegamimp > \tilomegad $.
The solute diffusion equation once again simplifies to Eq.\
\eqref{eq_semi_sol_diff_cpm}, where now $ \tileta $ involves $ \omegamimp $
instead of $ \omegarhoimp $.
With this change, and following a similar analysis as that described in the
inertial scenario, we once again arrive at Eq.\ \eqref{eq_semi_quadratic}.
The membrane frequency is then found to be given by
\begin{equation} \label{eq_semi_sol_membrane_omega}
	\omegam
	\, \approx \, \omegamimp \, \big(
		1
		\, \mp \, 8 \mk i \mk \Pm \mk \Os / \tileta
	\mk\big)
	~.
\end{equation}
Importantly, the sign of the imaginary part of $ \omegam $ is opposite that of
$ \omegarho $ [cf.\ Eq.\ \eqref{eq_semi_sol_inertia_omega}], and the out-of-%
phase drag force alters the inertial and membrane dynamics in different ways.
Substituting Eq.\ \eqref{eq_semi_sol_membrane_omega} into the diffusion equation
gives the concentration field [cf.\ Eq.\
\eqref{eq_semi_sol_inertia_hatcp}]
\begin{align}
	\hatcp
	\, &\sim \, \exp \mk \bigg\{
		\pm \, i \mk q \mk \tileta \mk z
		\, + \, \dfrac{4 \mk \Pm \mk \Os \mk q \mk z}{1 - \tilomegad / \omegamimp}
	\bigg\}
	\label{eq_semi_sol_membrane_hatcp}
	~.
\end{align}
The concentration field in Eq.\ \eqref{eq_semi_sol_membrane_hatcp} diverges as
$ z \rightarrow \infty $,
and is not physically meaningful.
As a consequence, the membrane branch vanishes when
$ q \notin (\qzm, \qzp) $.
A representative result at low tension is shown in Fig.\
\ref{fig_semi_diff_sols}, and at tensions larger than $ \lambdacast $
\eqref{eq_semi_lambdacast} the membrane branch vanishes entirely.

%
%
\medskip
\noindent\textbf{\textsf{The membrane branch when diffusion is fast.}}
When
$ \omegamimp < \tilomegad $
and
$ q \in (\qzm, \qzp) $,
the analysis of the coupled system is straightforward in the limit of small
$ \Pm $.
The diffusion equation, assuming
$ \omega = \omegamimp $
at lowest order, simplifies to
$ d^2 \hatcpm / d z^2 = (q^2 - \omegamimp / D) \mk \hatcpm $.
We thus find the exponentially decaying concentration solutions
\begin{equation} \label{eq_semi_sol_diff_fast_hatcpm}
	\hatcpm
	\, \sim \, \exp \mk \bigg\{
		\mp q \mk z \, \bigg(
			1
			\, - \, \dfrac{\omegamimp}{\tilomegad}
			\,
		\bigg)^{\!\! 1 \mkn / \mkn 2} \,
	\bigg\}
	~.
\end{equation}
With no oscillations in the lowest-order concentration field, modifications to
the hydrodynamic drag felt by the membrane are in phase with the membrane
velocity.
Moreover, as such corrections are of order
$ \Pm \! \ll \! 1 $,
we expect the membrane branch to be indistinguishable from its impermeable
counterpart---as confirmed in Fig.~\ref{fig_semi_diff_sols}.

%
%
\medskip
\noindent\textbf{\textsf{Theoretical implications.}}
As mentioned previously, solutes are present in the fluid surrounding nearly all
artificial and biological membranes.
Our findings demonstrate such systems can only be approximated as impermeable
for wavenumbers ``inside the dome,'' where
$ q \in (\qzm, \qzp) $
and there is a slow frequency
$ \omega \approx E / (4 \mk \muf \mk q) $.
In contrast, only a quickly decaying inertial mode exists for wavenumbers
$ q \notin (\qzm, \qzp) $
``outside the dome.''
For this case, ($i$) the membrane response is independent of $ \kb $ and
$ \lambdac $ at linear order, and ($i \mkn i$) thermal perturbations (discussed
subsequently) drive membrane undulations to continuously grow until
saturated by nonlinear forces---both of which are surprising results.

The predictions described above motivate us to question whether our normal mode
ansatz is indeed appopriate outside the dome.
Here, we examine this issue from a statistical mechanical perspective.
By starting with a microscopic Hamiltonian of the lipid, water, and solute
molecules, one can integrate over fast degrees of freedom to obtain a free
energy functional that depends only on the concentration field and membrane
height \cite{limmer}.
When solutes are fast relative to the membrane, concentration degrees of freedom
can also be integrated over to yield a free energy which depends only on the
membrane height.
We thus recover a description similar to that of the impermeable scenario.
If instead solutes are slow relative to the membrane, the concentration field is
not equilibrated for a given, instantaneous bilayer shape.
Upon integrating over the slow solute degrees of freedom, the resultant membrane
equation inerits this slowness via a memory kernel---for which the time
evolution of the membrane depends on its history, and there is no associated
free energy functional or equipartition result \cite{zwanzig}.
While such a result is incompatible with our normal mode ansatz, it justifies
the membrane behaving qualitatively differently inside and outside the dome.
Further analysis, likely involving nonlinear simulations of the membrane and
surrounding fluid \cite{narsimhan-jfm-2015, torres-jfm-2019, rower-jcp-2022,
tran-pre-2022, torres-pcb-2022, sahu-arxiv-2024, worthmuller-arxiv-2025}, is
required to understand the dynamics of the coupled system.
In any case, however, we emphasize that the standard impermeable result is not
applicable for wavenumbers outside the dome.

%
%

\section*{Experimentally characterizing GUVs}

In experimental investigations of lipid bilayers, one often seeks to
characterize membrane material properties prior to further manipulation.
When GUVs are involved, their fluctuations from thermal disturbances can be
imaged and---with a theory of membrane undulations---processed to determine
$ \kb $ and $ \lambdac $
\cite{pecreaux-epje-2004, ipsen-ch14, takatori-prl-2020, vutukuri-n-2020,
park-sm-2022, rautu-sm-2017, faizi-sm-2020, faizi-pnas-2024, lee-jpca-2020}.
While most experimental investigations apply the well-understood impermeable
theory over all wavenumbers, the present study reveals that only modes inside
the dome should be treated as impermeable.
Moreover, as the behavior of undulations outside the dome is not currently well%
-understood, experimental measurements involving these modes should be excluded
when determining membrane properties.
In what follows, we discuss how to incorporate thermal perturbations into our
theory, and then analyze GUV fluctuation data from experiments.

%
%
\medskip\noindent
\textbf{\textsf{The impermeable Langevin equation.}}
When the lipid bilayer is treated as impermeable, its linear response to
perturbations is well-known.
Since inertial modes decay much more quickly than their membrane counterparts
(see Eq.\ \eqref{eq_imp_height_branches} with $ \omegarho \gg \omegam $),
inertia can be neglected entirely; the membrane frequency $ \tilomegam $
presented in Eq.\ \eqref{eq_imp_tilomega} is then exact.
Decomposing the membrane height as
$
	h(x, y, t)
	\, = \, \sum_{\bmq} \tilhq (t) \mk e^{i \mk (\qx x \, + \, \qy y) }
$
leads to the Langevin equation
[cf.\ Eqs.\ \eqref{eq_imp_height_fourier},\eqref{eq_imp_quadratic}]
\begin{equation} \label{eq_exp_langevin}
	\dd{\tilhq}{t}
	\, + \, \tilomegam \mk \tilhq
	\, = \, \tilxiq (t)
	~.
\end{equation}
In Eq.\ \eqref{eq_exp_langevin}, $ \tilxiq (t) $ is a complex Gaussian thermal
noise satisfying the fluctuation--dissipation theorem, with two-time
correlations proportional to
$ \delta (t - t') \, \kBT / (4 \mk \muf \mk q \, \lc^{\, 2}) $
\cite{sapp-pre-2016}.
Equation \eqref{eq_exp_langevin} can be solved exactly to yield
\cite{chandrasekhar-rmp-1943}
\begin{equation} \label{eq_exp_equipartition}
	\hqsq
	\, = \, \dfrac{\kBT}{E \mk \lc^{\, 2}}
	~,
\end{equation}
which is the well-known result of the equipartition theorem as applied to planar
membranes \cite{brochard-jp-1975}.

%
%
\medskip\noindent
\textbf{\textsf{The Langevin equation inside the dome.}}
For modes with
$ q \in (\qzm, \qzp) $,
there is no discernible difference between the impermeable and semipermeable
dynamics.
Thus, Eq.\ \eqref{eq_exp_langevin} once again applies, and membrane fluctuations
satisfy the equipartition result \eqref{eq_exp_equipartition}.

%
%
\medskip\noindent
\textbf{\textsf{The Langevin equation outside the dome.}}
As discussed previously, we are currently unsure if our normal mode ansatz is
appropriate outside the dome.
If the ansatz is indeed valid, then we effectively see only an inertial mode.
The corresponding Langevin equation describing membrane fluctuations is given by
\begin{equation} \label{eq_exp_langevin_inertia}
	\dfrac{1}{\omegarhor} \, \ddt{\tilhq}{t}
	\, + \, \dd{\tilhq}{t}
	\, = \, \tilxiq (t)
	~,
\end{equation}
where $ \omegarhor $ is the real part of the frequency and $ \tilxiq $ is
the same thermal noise as in Eq.\ \eqref{eq_exp_langevin}.
There is no elastic restoring force in Eq.\ \eqref{eq_exp_langevin_inertia},
which is analogous to the equation of motion for a colloidal particle in a
fluid.
Just as a colloid diffuses freely at long times \cite{chandrasekhar-rmp-1943},
the variance of membrane height undulations grows linearly in time according to
Eq.\ \eqref{eq_exp_langevin_inertia}.
Eventually, the membrane height will grow to a point where our assumption of
linearity is no longer valid.
At this point, membrane bending and surface tension forces will presumably
prevent height fluctuations from continuing to grow---yet a description of such
behavior is beyond the scope of the present work.
Thus, irrespective of whether our ansatz is valid, membrane fluctuations are not
expected to satisfy the equipartition result outside the dome where
$ q \notin (\qzm, \qzp) $.

\begin{figure*}[b!]
	\centering\vspace{-14pt}
	\includegraphics[width=\textwidth]{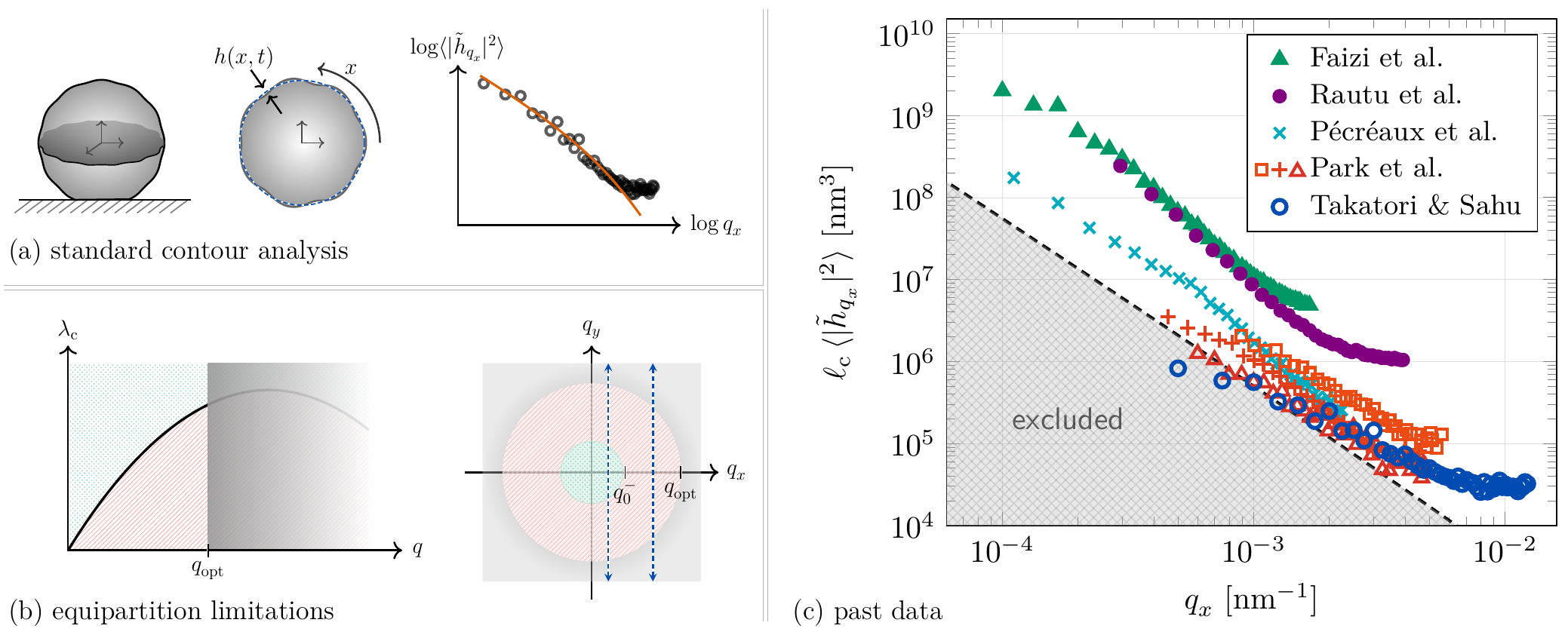}\vspace{-5pt}
	\caption{%
		Analysis of GUV contour fluctuations, modified due to membrane
		semipermeability.
		(a) Canonical protocol to calculate the bending modulus and surface tension
		of a fluctuating GUV.
		In a spherical vesicle of radius $ \rv $ (left), the equatorial
		cross-section is imaged (center) at successive time intervals.
		Radial deviations from a circle are recorded as $ h(x, t) $, where $ x $ is
		the arclength along the unperturbed circle.
		The shape disturbances $ h(x, t) $ are decomposed into one-dimensional
		Fourier modes, whose thermally-averaged amplitudes (right; adapted from
		Ref.\ \cite{takatori-prl-2020}) are fit to Eq.\ \eqref{eq_exp_hqxsq} by
		tuning the values of $ \kb $ and $ \lambdac $ (right; solid orange line)
		\cite{pecreaux-epje-2004}.
		Theory and experiment deviate at large $ \qx $ due to finite optical
		resolution in the latter.
		(b) Limitations of the equipartition result \eqref{eq_exp_equipartition} due
		to solutes in the surrounding fluid.
		(left) Modes inside the dome (red lines) satisfy Eq.\
		\eqref{eq_exp_equipartition}; those outside the dome (green dots) do not.
		Membrane undulations cannot be visualized above 
		$ \qopt \approx 0.025 $ nm$ ^{-1} $
		(gray shading) due to limitations in the optical resolution
		\cite{ipsen-ch14}.
		(right) For a GUV with surface tension
		$ \lambdac \in (0, \lambdacast) $,
		modes are characterized in the $ \qx $--$ \qy $ plane.
		The equipartition result \eqref{eq_exp_equipartition} is limited to modes
		with amplitude
		$ q > \qzm $,
		and large-$ q $ modes cannot be visualized.
		The domains of integration in Eq.\ \eqref{eq_exp_hqxsq}, corresponding to
		two different values of $ \qx $, are shown as dashed vertical lines.
		The result of Eq.\ \eqref{eq_exp_hqxsq} is invalid when
		$ \qx < \qzm $,
		as in such cases the integral contains modes outside the dome.
		(c) Plot of experimental membrane fluctuations, over a range of reported
		surface tensions $ \lambdac $, from prior studies:
		Faizi et al.\ \cite{faizi-sm-2020}
		($ 3.1 \! \cdot \! 10^{-6} $ pN$/$nm),
		Rautu et al.\ \cite{rautu-sm-2017}
		($ 1.4 \! \cdot \! 10^{-5} $ pN$/$nm),
		P\'ecr\'eaux et al.\ \cite{pecreaux-epje-2004}
		($ 1.7 \! \cdot \! 10^{-4} $ pN$/$nm),
		Park et al.\ \cite{park-sm-2022}
		($ 7 \! \cdot \! 10^{-4} $--$ 3 \! \cdot \! 10^{-3} $ pN$/$nm),
		and
		Takatori \& Sahu \cite{takatori-prl-2020}
		($ 4 \! \cdot \! 10^{-3} $ pN$/$nm).
		The $ y $-axis is chosen to be independent of the vesicle radius, and
		depends only on $ \lambdac $ and $ \kb $ [see Eq.\ \eqref{eq_exp_hqxsq}].
		At large $ \qx $, the data saturates due to optical resolution limitations.
		All data points below the dashed black line are outside the dome, and should
		be excluded when determining membrane parameters.
		Fluctuation data from the high-tension vesicle in Ref.\
		\cite{vutukuri-n-2020} is not publicly available, and so cannot be plotted%
		---though the first 88 modes should be excluded based on the reported value
		of
		$ \lambdac = 0.025 $ pN$/$nm
		\cite[\S4]{supplemental}.
	}
	\label{fig_exp}
\end{figure*}

%
%
\medskip\noindent
\textbf{\textsf{The standard contour analysis of a GUV.}}
Figure \figpart{fig_exp}{a} portrays the canonical analysis of a GUV undergoing
thermal fluctuations, the details of which can be found in Ref.\
\cite{pecreaux-epje-2004}.
In short, the equatorial cross-section of a GUV is imaged at frequent time
intervals.
For each snapshot, the radial deviation from a circle of radius $ \rv $ is
measured and decomposed into normal modes.
As GUVs are large and weakly curved, it is often convenient to apply the planar
membrane theory \cite{pecreaux-epje-2004}---which is simpler than its spherical
counterpart \cite{rautu-sm-2017, faizi-sm-2020, faizi-pnas-2024, ipsen-ch14}.
The position $ x $ is defined as the distance along the circle, with $ h(x, t) $
the radial perturbation,
$ \lc = 2 \mk \pi \mk \rv $,
and
$ \qx = 2 \mk \pi \mk m / \lc = m / \rv $
for positive integers $ m $.
The one-dimensional (1D) fluctuation spectrum $ \hqxsq $ is calculated from
experimental measurements as in Fig.\ \figpart{fig_exp}{a}.
To compare the 1D data with the theoretical 2D equipartition result of Eq.\
\eqref{eq_exp_equipartition}, the latter is averaged over all $ \qy $ as
\cite{pecreaux-epje-2004}
\begin{equation} \label{eq_exp_hqxsq}
	\begin{split}
		\hqxsq
		\,&= \, \dfrac{\lc}{2 \mk \pi} \int_{- \infty}^\infty \md \qy \, \hqsq
		\\[1pt]
		\,&= \, \dfrac{\kBT}{2 \mk \lambdac \mk \lc} \, \bigg(
			\dfrac{1}{\qx}
			\, - \, \dfrac{1}{\sqrtl{\qxsq + 2 \mk \lambdac / \kb} \,}
		\bigg)
		~.
	\end{split}
\end{equation}
The bending modulus $ \kb $ and surface tension $ \lambdac $ are chosen so as to
minimize the difference between the measured fluctuation spectrum and prediction
of Eq.\ \eqref{eq_exp_hqxsq}---which, in Fig.\ \figpart{fig_exp}{a}, are
respectively shown as black circles and an orange line.
Note that short-wavelength fluctuations cannot be visualized in experiments due
to a lower bound
$ \lopt \approx 250 $ nm
on the optical resolution, which sets an upper limit
$ \qopt := 2 \mk \pi / \lopt \approx 0.025 $ nm$ ^{-1} $
beyond which undulations cannot be observed \cite{ipsen-ch14}.
In practice, short-wavelength modes are excluded when fitting $ \kb $ and
$ \lambdac $.
Techniques also exist to correct for the finite camera integration time
\cite{pecreaux-epje-2004} and vertical focal projections \cite{rautu-sm-2017},
but are not discussed here.

%
%
\medskip\noindent
\textbf{\textsf{The modified contour analysis of a GUV.}}
When solutes are present in the fluid surrounding the membrane, the 2D
equipartition result \eqref{eq_exp_equipartition} is only valid for modes
$ \bmq $ with magnitude
$ \lvert \bmq \rvert \in (\qzm, \qzp) $.
We seek to determine how the experimental analysis of contour fluctuations is
altered by such a restriction.
To this end, we first note that
$ \qopt < \qzp $, as shown in Fig.\ \figpart{fig_exp}{b}.
Short-wavelength undulations outside the dome thus cannot be observed in
experiments, and do not contribute to the determination of $ \kb $ and
$ \lambdac $.
Long-wavelength fluctuations outside the dome, on the other hand, are captured
experimentally.
Figure \figpart{fig_exp}{b} characterizes modes in the $ \qx $--$ \qy $ plane as
being either outside the dome
($ \lvert \bmq \rvert < \qzm $),
inside the dome and visually accessible
($ \qzm < \lvert \bmq \rvert < \qopt $),
or visually inaccessible
($ \lvert \bmq \rvert > \qopt $).
Here, domains of the contour integral in Eq.\ \eqref{eq_exp_hqxsq} for two
values of $ \qx $ are shown as dashed vertical lines.
If
$ \qx < \qzm $,
this domain contains modes for which the equipartition result
\eqref{eq_exp_equipartition} does not apply.
The result of Eq.\ \eqref{eq_exp_hqxsq}, which relies on Eq.\
\eqref{eq_exp_equipartition}, is accordingly valid only when
$ \qx > \qzm $%
---with the understanding that as $ \qx $ increases, fewer modes are optically
accessible.

To determine $ \kb $ and $ \lambdac $ from GUV contour fluctuations, it is
useful to plot fluctuation amplitudes as in Fig.\ \figpart{fig_exp}{c}---which
contains experimental data from several prior investigations
\cite{faizi-sm-2020, rautu-sm-2017, pecreaux-epje-2004, park-sm-2022,
takatori-prl-2020}.
Here, both the $ x $-axis and $ y $-axis are chosen to be independent of $ \rv $
[cf.\ Eq.\ \eqref{eq_exp_hqxsq}].
Next, experimental data inside and outside the dome need to be identified.
To this end, we combine Eqs.\ \eqref{eq_semi_qzpm} and \eqref{eq_exp_hqxsq} to
obtain the fluctuation amplitudes of 1D modes
$ \qx = \qzm $
on the dome, for tensions
$ \lambdac \in (0, \lambdacast) $:
\begin{equation} \label{eq_exp_dome_plot}
	\lc \mk \hqxsq
	\, = \ \dfrac{
		\kBT \, (1 - \sqrtl{\kb \mk \qx / (8 \mk \muf D)} \,)
	}{
		\qxsq \mk (8 \mk \muf D - \kb \mk \qx)
	}
	~.
\end{equation}
Importantly, $ \kb $ is not known a priori, and so Eq.\ \eqref{eq_exp_dome_plot}
can be approximated as
$ \lc \mk \hqxsq \approx \kBT / (8 \mk \muf \mk \qxsq D) $%
---valid when
$ \kb \mk \qx \ll 8 \mk \muf \mk D $%
---in a first analysis of the data.
Equation \eqref{eq_exp_dome_plot} is plotted as the dashed black line in Fig.\
\figpart{fig_exp}{c}; all of the data points below it are outside the dome and
should be excluded when determining $ \kb $ and $ \lambdac $.

Our aggregation and analysis of the experimental data in Fig.\
\figpart{fig_exp}{c} is detailed in \S4 of the SM \cite{supplemental}.
Most often, vesicle tensions
$ \lambdac \! \sim \! 10^{-7} $--$ 10^{-4} $ pN$/$nm
are low \cite{pecreaux-epje-2004, rautu-sm-2017, faizi-sm-2020, faizi-pnas-2024}
and all data lie inside the dome.
As it turns out, a vesicle at low tension is experimentally favorable because
its undulations are large and hence easy to image.
In many cases, the vesicle's osmotic environment is manipulated to elicit a low
tension.
Notable exceptions are the so-called active vesicles containing either self-%
propelled Janus particles or motile bacteria
\cite{takatori-prl-2020, vutukuri-n-2020, park-sm-2022}.
For these systems, moderate vesicle tensions are required for active particles
to not overcome the elastic membrane restoring force and form a tube.
Given the fluctuation data for active vesicles in Fig.\ 1 of Ref.\
\cite{takatori-prl-2020} and Fig.\ 6(c) of Ref.\ \cite{park-sm-2022}, between
one and eight long-wavelength modes lie outside the dome
\cite[\S4]{supplemental}---though the overall findings of these studies are
unaffected.
In contrast, for the high-tension
($ \lambdac = 0.025 $ pN$/$nm)
active vesicle in Fig.\ 1(g) of Ref.\ \cite{vutukuri-n-2020}, the first 88
long-wavelength modes lie outside the dome.
It is likely that upon excluding these modes, the calculated values of $ \kb $
and $ \lambdac $ would be affected.
Unfortunately, the experimental fluctuation data for this vesicle is not
provided, and we cannot investigate further.
Nevertheless, we find osmotic forces can indeed alter membrane fluctuations in
biologically-relevant scenarios.

%
%

\section*{Concluding remarks}

In this work, we investigated the linearized dynamics of a planar lipid
membrane, surrounded by a Newtonian fluid in which solutes are dissolved.
Though the lipid bilayer is only weakly permeable to water, the presence of
solutes can significantly alter its dynamics.
More specifically, when
$ \tilomegad < \tilomegam $
and diffusion is slow relative to the impermeable membrane, undulations of the
semipermeable system no longer decay at a frequency near $ \tilomegam $.
In such cases, either ($i$) the ansatz that all quantities can be expressed in
terms of planar normal modes as in Eqs.\ \eqref{eq_imp_height_fourier},
\eqref{eq_imp_pressure_fourier}, and \eqref{eq_semi_concentration_fourier} is no
longer valid, or ($i\mkn i$) the bilayer only decays at the fast inertial
frequency
$ \omegarho \gg \omegam $.
When thermal perturbations are incorporated into the theory, the well-known
equipartition result \eqref{eq_exp_equipartition} is found to only be
experimentally relevant when
$ q \in (\qzm, \qopt) $%
---a range that narrows as $ \lambdac $ is increased.
Our results are relevant when membrane tensions are large, as is the case with
active vesicles \cite{takatori-prl-2020, vutukuri-n-2020, park-sm-2022} and a
variety of biological scenarios
\cite{dai-bpj-1995, dai-jn-1998, datar-bpj-2019, pullarkat-prl-2006,
tchoufag-prl-2022, goldstein-jp-1996, powers-prl-1997, boulant-ncb-2011,
al-izzi-prl-2018, shi-cell-2018, al-izzi-arxiv-2024, janssen-pre-2024,
venkatesh-jfm-2025}.

The vanishing of the membrane mode at low $ q $ and large $ \lambdac $ opens
several lines of inquiry.
We are particularly interested to see if membranes in spherical or cylindrical
geometries respond in the same fashion; the former is needed for a careful
analysis of GUV undulations \cite{rautu-sm-2017, faizi-sm-2020, faizi-pnas-2024}
while the latter could be relevant to neuronal systems \cite{dai-bpj-1995,
dai-jn-1998, pullarkat-prl-2006, datar-bpj-2019, tchoufag-prl-2022}.
In all scenarios where the membrane branch vanishes, the behavior of the coupled
membrane--fluid--solute system can be interrogated with nonlinear simulations
\cite{narsimhan-jfm-2015, torres-jfm-2019, rower-jcp-2022, tran-pre-2022,
torres-pcb-2022, sahu-arxiv-2024, worthmuller-arxiv-2025}.
Additional levels of complexity can be introduced by considering membranes that
are not ideally selective \cite{alkadri-pre-2025}, have finite thickness
\cite{lipel-prf-2025, omar-pre-2024-i, omar-pre-2025-ii, omar-pre-2025-iii}, or
are surrounded by charged solutes---with the latter requiring extensive
theoretical developments \cite{fong-aiche-2020, omar-pre-2024-i,
omar-pre-2025-ii, omar-pre-2025-iii, yu-pre-2025}.

%
%
\medskip\small\balance
\noindent\textbf{\textsf{Data, methods, and software availability.}}
The Supplemental Material \cite{supplemental} details the theoretical techniques
used, as well as the methods of numerical solution.
The code used to generate all data in the present study is publicly available at
\href{https://github.com/sahu-lab/osmosis-flat}%
{\texttt{github.com/sahu-lab/osmosis-flat}}.

\small

%
%
\medskip
\noindent\textbf{\textsf{Acknowledgments.}}
We express our most sincere gratitude to Prof.\
\href{https://www.cchem.berkeley.edu/~kranthi/}{Kranthi Mandadapu}
for his perspective on lipid membrane behavior.
We are also thankful to have discussed Ref.\ \cite{alkadri-pre-2025} with Dr.\
\href{https://www.researchgate.net/profile/Ahmad-M-Alkadri}{Ahmad Alkadri}
and Kranthi, as well as Refs.\ \cite{lee-jpca-2020,park-sm-2022} with Dr.\
Kisung Lee, Dr.\ Myoenggon Park, and Prof.\
\href{https://www.umass.edu/polymer-science/about/directory/steve-granick}{Steve Granick}.
It is a pleasure to thank Dr.\
\href{https://scholar.google.com/citations?hl=en&user=UrwMUscAAAAJ&view_op=list_works&sortby=pubdate}{Jo\"el Tchoufag}
for many stimulating discussions,
Profs.\
\href{https://physics-astronomy.jhu.edu/directory/brian-camley/}{Brian Camley}
and
\href{https://engineering.jhu.edu/faculty/tine-curk/}{Tine Curk}
regarding their thoughts on membrane fluctuations, and Prof.\
\href{https://cheme.stanford.edu/people/sho-takatori-1}{Sho Takatori} for
introducing us to the experimental analysis of GUV fluctuations.
This work was partially supported by the \href{https://welch1.org/}{Welch
Foundation} via Grant No.\ F-2208-20240404.

%
%
\medskip\small
\noindent\textbf{\textsf{Declaration of interests.}}
There are no conflicts to declare.

\small%
\subsection*{References}
\bibliographystyle{bibStyle}
\bibliography{refs}

\end{document}